\documentclass[preprints,article,accept,oneauthor]{Definitions/mdpi}
\usepackage{xcolor}
\usepackage{comment}
\firstpage{1}
\pubvolume{1}
\issuenum{1}
\articlenumber{0}
\pubyear{2026}
\copyrightyear{2026}
\externaleditor{Firstname Lastname}
\datereceived{}
\daterevised{}
\dateaccepted{}
\datepublished{}

\newcommand{\dd}{\mathrm{d}}

\newcommand{\ii}{\mathrm{i}}
\newcommand{\pp}{p_{\perp}}
\newcommand{\pa}{p_{\parallel}}
\newcommand{\ompe}{\omega_{pe}}
\newcommand{\kb}{\kappa}
\newcommand{\epsz}{\varepsilon_0}

\newcommand{\bfk}{\boldsymbol{k}}
\newcommand{\bfv}{\boldsymbol{v}}
\newcommand{\bfp}{\boldsymbol{p}}
\newcommand{\nablaP}{\boldsymbol{\nabla}_{p}}

\Title{Linear Electrostatic Instabilities Driven by Relativistic Non-Maxwellian Pair Beams}

\Author{Suman %MDPI: Please carefully check the accuracy of names and affiliations. % Suman Dey: Checked and verified.
 Dey \orcidA{}}
\AuthorNames{Suman Dey}

\address[1]{
  Institut f\"{u}r Theoretische Physik, Technische Universit\"{a}t Braunschweig, 38106 Braunschweig, Germany; s.dey@tu-bs.de %MDPI: The email addresses highlighted are different from the ones submitted online at susy.mdpi.com. Please confirm which are correct.
}

\abstract{We investigate how non-Maxwellian momentum distributions affect the linear growth of electrostatic instabilities driven by relativistic pair beams, motivated by the lack of a universally accepted analytical description of blazar-induced pair beams. Different beam distributions are used to describe relativistic pair beams in analytical studies and particle-in-cell simulations. A systematic comparison is, therefore, needed to determine how the beam distribution influences the instability growth rates. We formulate the relativistic electrostatic response in the weak-beam limit for an anisotropic $\kappa$-type distribution and compare the resulting growth rates with those obtained for Gaussian and Maxwell--J\"{u}ttner distributions. We require $\kappa>3/2$ for a finite root-mean-square momentum spread. For the parameters considered, the $\kappa$-type distributions yield higher maximum growth rates than the Gaussian distribution, even when their mean momenta and momentum spreads are the same. The fastest-growing modes correspond to parallel propagation for the $\kappa$ and Gaussian distributions, whereas the Maxwell--J\"{u}ttner distribution exhibits its maximum for oblique propagation for a mildly relativistic beam. An additional calculation for a highly relativistic pair beam with $\kappa=2$ finds an oblique maximum of the electrostatic growth rate. The beam momentum distribution influences the linear growth of electrostatic instabilities under the conditions investigated.}

\keyword{beam--plasma instabilities; plasma kinetic theory; electrostatic waves and oscillations; astrophysical plasmas; blazars}

\begin{document}

\section{Introduction}

Very-high-energy (VHE) gamma rays emitted by blazars interact with the extragalactic background light (EBL) \cite{Hauser:2001xs} during their propagation through intergalactic space, producing relativistic electron--positron pairs \cite{Aharonian:1993vz,Berezinsky:2016feh}. These pairs can further upscatter cosmic microwave background \cite{gould1967opacity, blumenthal1970bremsstrahlung} and EBL photons through inverse-Compton scattering, giving rise to secondary gamma-ray emission in the GeV energy range. However, this expected cascade emission is absent in the observations of some blazars. One possible explanation is the deflection of the electron--positron pairs by weak intergalactic magnetic fields. Another possibility is that the pairs lose part of their energy through collective plasma processes before the cascade develops fully. In particular, beam--plasma instabilities have been studied as a possible energy-loss mechanism for blazar-induced pair beams, although their overall importance remains debated in the literature \cite{PerryLyubarsky2021, AlawashraPohl2022, AlawashraPohl2024,AlawashraPohl2026, Dey:2025wdj, Dey:2026nml}. As the beams produced from pair production interactions propagate through the intergalactic medium (IGM), they can excite collective plasma waves through beam--plasma instabilities. The linear growth of these electrostatic and electromagnetic modes has been studied using relativistic kinetic theory and numerical simulations \cite{fainberg1970nonlinear,bret2010multidimensional,SchlickeiserKrakauSupsar2013,SchlickeiserIbscherSupsar2012,Rafighi:2017ise,Vafin:2018kox}. The beam--plasma instability can excite both electrostatic and electromagnetic modes. However, for blazar-induced pair beams, electrostatic oblique modes generally dominate over electromagnetic Weibel/filamentation-type modes, because the beam shows momentum anisotropy with a large longitudinal momentum spread compared with its transverse momentum \cite{Dey:2025wdj, Rafighi:2017ise, AlawashraPohl2022}.

Since the electrostatic growth rate is determined by the particles satisfying the wave--particle resonance condition, it depends on the momentum gradients of the beam distribution. The distribution function is, therefore, an important ingredient in any linear analysis of the instability. The momentum distribution of blazar-induced pair beams needs to be better understood. The beam momentum distribution depends on the primary gamma-ray spectrum, the pair-production process, and the subsequent propagation of the resulting particles. In previous studies, different approaches have adopted different distributions, depending on the physical question and the numerical method. For example, Rafighi et al. \cite{Rafighi:2017ise} and Vafin et al. \cite{Vafin:2018kox} considered a Maxwell--J\"{u}ttner type pair distribution in their numerical analyses. Beck et al. \cite{BeckEtAl2023} studied the evolution of a pair beam with a Gaussian momentum distribution, while our previous numerical study \cite{Dey:2025wdj} used a Cauchy-type distribution to represent the non-Maxwellian nature of the beam. They raise an important question: how does the beam distribution affect the growth of plasma instabilities when the other beam parameters are kept fixed? The linear growth rate provides information about the initial development of an instability, whereas the subsequent evolution depends on the nonlinear interaction between the beam and the excited plasma waves. Therefore, an analytical description of the linear stage is useful for comparing different distribution functions and identifying growth rates before turning to more expensive kinetic simulations.

Previous work has already established that the beam momentum distribution influences electrostatic instability growth. In particular, Perry and Lyubarsky~\cite{PerryLyubarsky2021} and Alawashra and Pohl~\cite{AlawashraPohl2024} examined resonant instabilities of exceptionally narrow blazar-induced pair beams and distinguished angular broadening from efficient beam-energy transfer in their quasi-linear treatment. The present investigation instead addresses a controlled linear-response question: how varying the tails of an anisotropic $\kappa$-type distribution changes the resonant growth rates and their wavevector dependence relative to Gaussian and Maxwell--J\"{u}ttner reference distributions. Our analysis focuses on the influence of different momentum distributions on linear instability growth while keeping the beam parameters fixed, without considering the time evolution of the cascade or nonlinear relaxation processes.

Observations of gamma-ray emission from blazars also raise important questions about the evolution of relativistic pair beams. For example, Fermi-LAT observations and searches for extended gamma-ray emission around BL Lac objects have found no significant evidence of pair halos in the investigated sources, leading to upper limits on their emission \cite{Ackermann2013PairHalos}. However, these constraints depend on the past gamma-ray activity of the sources, intergalactic magnetic field strengths, and cascade propagation, and, therefore, cannot be used to directly determine the growth rates of plasma instabilities.

The $\kb$-type distribution allows the strength of the high-momentum tail to be controlled by $\kb$, while separate momentum-scale parameters determine the spreads parallel and perpendicular to the beam direction. As $\kb$ increases, the distribution approaches a Gaussian. Relativistic $\kb$-distributions have been studied in other plasma contexts, and methods have been developed to initialize them in particle-in-cell simulations \cite{Zenitani:2022pmm,Mousavi:2024waq}. However, the $\kb$-distribution used in previous studies represents a specific momentum-space model and may not describe all relativistic pair beams produced by blazar cascades. The purpose of this work is to derive a consistent linear-response framework for an anisotropic, $\kb$-type distribution and to identify the exact integrals that determine the growth of electrostatic oblique modes. We focus on the weak-beam regime, in which the beam density is much lower than the background plasma density, consistent with blazar-induced beams.

The paper is organized as follows. In Section~\ref{sec:beam-dist}, we introduce an anisotropic $\kb$-type momentum distribution and derive its normalization. We also describe the Gaussian and Maxwell--J\"{u}ttner distributions used for comparison. In Section~\ref{sec:dispersion}, we derive the linear electrostatic dispersion relation and obtain the growth rate in the weak-beam limit, reducing the resonant integral to a two-dimensional integral by performing the gyrophase integration analytically. In Section~\ref{sec:growth}, we present the numerical growth-rate maps and compare the maximum growth rates and corresponding wavevectors for the different beam distributions. In Section
\ref{sec:astrophysical}, we extend the analysis to a highly relativistic pair beam and examine its electrostatic instability growth rate and preferred propagation direction. Finally, Section~\ref{sec:conclusion} summarizes the main results and discusses the limitations of the present analysis and directions for future work.
%MDPI: Please confirm if the bold formatting is necessary; if not, please remove it. %Suman Dey: Corrected.

\section{Physical Model and Beam Distribution}\label{sec:beam-dist}

%\subsection{Relativistic kinematics}
We consider a spatially homogeneous, unmagnetized background plasma representative of the intergalactic medium through which blazar-induced relativistic electron–positron pairs propagate. We denote the number density of the background plasma electrons by $n_p$ and the electron mass by $m_e$. The characteristic frequency of collective electron oscillations, known as the electron plasma frequency, is given by \cite{chen2018introduction, stix1992waves} %MDPI: Please ensure all variables/values in the equation appear in the same format in the text (normal/italic/bold/subscript/superscript). %Suman Dey: all good.
\begin{equation}
\omega_{pe} = \left(\frac{n_p e^2}{\epsilon_0 m_e}\right)^{1/2},
\label{eq:w_pe}
\end{equation}
where $e$ is the charge of electron and $\epsilon_0$ is the vacuum permittivity. In the blazar-induced pair-beam scenario, $\omega_{pe}$ provides a characteristic frequency scale for the plasma waves excited by the relativistic pair beam. We assume the ions remain effectively stationary on the electron-plasma timescale. The number density of the beam is defined by $n_b$, and the beam propagates along the $z$-axis. We consider a relativistic electron–positron beam in which both species share the same momentum distribution. The total beam number density, $n_b$, is defined as the sum of the electron and positron number densities. Although electrons and positrons have opposite charges, their contributions to the linear susceptibility do not cancel. This is because the susceptibility depends on the square of the particle charge, so both species contribute equally when they follow the same momentum distribution.

 {For a homogeneous, fully ionized intergalactic background with fixed ionization fraction and negligible overdensity, the electron number density approximately follows $n_p(z)=n_{p,0}(1+z)^3$. Equation~\eqref{eq:w_pe}, therefore, implies $\omega_{pe}(z)=\omega_{pe,0}(1+z)^{3/2}$. This relation describes the redshift evolution of the background plasma frequency, whereas determining how the beam density and momentum distribution evolve with redshift requires a separate physical model.}

The transverse and longitudinal components of the particle momentum are defined as $\pp=(p_x^2+p_y^2)^{1/2}$ and $\pa=p_z$, respectively, taking the $z$-axis to be the direction of beam propagation. The relativistic Lorentz factor and particle velocity are then given by
\begin{equation}
  \gamma(\bfp)=\left[1+\frac{\pp^2+\pa^2}{m_e^2c^2}\right]^{1/2},
  \qquad \bfv=\frac{\bfp}{\gamma m_e},
  \label{eq:relkin}
\end{equation}
where $\bfp=(p_x,p_y,p_z)$ is the three-dimensional momentum vector of every beam particle and $c$ is the speed of light. Since the beam distribution is symmetric about the $z$-axis, the wavevector can be written as $\bfk=(k_\perp,0,k_\parallel)$, with its transverse component along the $x$-direction. Here, $k_\perp$ and $k_\parallel$ are the components perpendicular and parallel to the beam axis, respectively, and $k=(k_\perp^2+k_\parallel^2)^{1/2}$ is the magnitude of the wavevector.
%The propagation angle measured from the beam axis is \(\theta_k=\arctan(k_\perp/k_\parallel)\), with the appropriate quadrant understood.

\subsection{Anisotropic \texorpdfstring{$\kappa$}{kappa}-Type Distribution}

In the blazar-induced pair-beam scenario, high-energy gamma rays interact with the extragalactic background light and produce relativistic electron--positron pairs. These pairs propagate predominantly along the direction of the primary gamma rays. However, the beam has finite angular and momentum spreads, meaning that the particles do not all move in exactly the same direction or carry the same momentum. A monoenergetic, perfectly collimated beam, therefore, lacks some physical structure. A more realistic momentum distribution is needed to investigate how these finite spreads affect the beam--plasma instability. Previous studies \cite{Vafin:2018kox, Rafighi:2017ise}, in this context, have used a Maxwell-J\"{u}ttner-type distribution to describe relativistic beams. In this work, we introduce an anisotropic, shifted $\kb$-type distribution to account for different momentum spreads parallel and perpendicular to the beam direction and to capture non-Maxwellian features. We then compare the resulting instability growth rates with those obtained using Gaussian and Maxwell--J\"{u}ttner distributions. The choice of a $\kb$-type distribution is motivated by its ability to describe populations with extended, power-law-like tails, which cannot be represented adequately by a Gaussian or a Maxwell--J\"{u}ttner distribution. 

We define the beam distribution by
%However, to investigate how different momentum spreads parallel and perpendicular to the beam direction, as well as possible non-Maxwellian features, we adopt an anisotropic, shifted $\kb$-type distribution to study the beam--plasma instability,
\begin{equation}
 f_{\kb}(\pp,\pa)=C_{\kb}
 \left[1+\frac{\pp^2}{\kb a_\perp^2}
          +\frac{(\pa-p_0)^2}{\kb a_\parallel^2}
          \right]^{-(\kb+1)},
 \label{eq:kappadist}
\end{equation}
where $p_0>0$ is the mean longitudinal momentum, and $a_\perp$ and $a_\parallel$ are the characteristic transverse and longitudinal momentum scales, respectively. The parameter $\kb$ determines the shape of the distribution, particularly the relative importance of its tails. Smaller values of $\kb$ produce more extended power-law tails, whereas the distribution approaches an anisotropic Gaussian as $\kb\to\infty$. The different transverse and longitudinal scales allow the momentum distribution to define the anisotropy expected for a directed relativistic beam. The normalization constant $C_{\kb}$ is determined by requiring the integral of the distribution over momentum space to equal the total beam number density $n_b$. In cylindrical momentum coordinates, the integration can be expressed as
\begin{equation}
  \dd^3p=2\pi\pp\,\dd\pp\,\dd\pa,
  \qquad \int f_{\kb}(\bfp)\,\dd^3p=n_b.
  \label{eq:measure}
\end{equation}

%MDPI: We added indent, please confirm, same below. % Suman Dey: all right.
We further introduce the dimensionless variables $u=\pp/a_\perp$ and $s=(\pa-p_0)/a_\parallel$. Integrating over the azimuthal angle reduces the normalization integral to a standard radial integral in three-dimensional momentum space. In particular, the identity
\begin{equation}
 \int_{\mathbb{R}^3}\left(1+\frac{q^2}{\kb}\right)^{-(\kb+1)}\dd^3q
 =\pi^{3/2}\kb^{3/2}\frac{\Gamma(\kb-1/2)}{\Gamma(\kb+1)},
 \label{eq:betaintegral}
\end{equation}
gives
\begin{equation}
 C_{\kb}=\frac{n_b\,\Gamma(\kb+1)}
{\pi^{3/2}\kb^{3/2}a_\perp^2a_\parallel\,\Gamma(\kb-1/2)}.
 \label{eq:norm}
\end{equation}

Here, $\Gamma(\cdot)$ denotes the Euler gamma function.
The integral converges for $\kb>1/2$. The first absolute momentum moments require $\kb>1$, and finite second moments require $\kb>3/2$. 
In our previous numerical study \cite{Dey:2026ecu,Dey:2025wdj}, we used a Cauchy-like distribution corresponding to $\kb=1$ to model the non-Maxwellian momentum distribution of the beam. However, for the three-dimensional $\kb$-type distribution defined here, the existence of finite momentum imposes a restriction on $\kb$. In particular, although the distribution is normalizable for $\kb>1/2$, its second momentum moments converge only for $\kb>3/2$. Consequently, the root-mean-square momentum spreads parallel and perpendicular to the beam direction are well defined only for $\kb>3/2$. Since the present study aims to compare different distribution functions at fixed momentum spreads, we restrict our analysis to $\kb>3/2$. Thus, the $\kb$-type distribution considered here provides a systematic way to investigate the influence of non-Maxwellian tails on the linear beam--plasma instability.

\subsection{Properties of the \texorpdfstring{$\kappa$}{kappa}-Distribution and the Gaussian Limit}\label{subsec:Gaussian-limit}

By symmetry, the distribution is centered at $p_\parallel=p_0$ and has no preferred direction in the transverse plane. Therefore, the mean momentum components are
\begin{equation}
 \langle p_\parallel\rangle=p_0,
 \qquad \langle p_x\rangle=\langle p_y\rangle=0.
 \label{eq:means}
\end{equation}

For $\kb>3/2$, the variances of the momentum components are finite, allowing us to define the root-mean-square momentum spreads parallel and perpendicular to the beam direction. The corresponding variances are
\begin{equation}
 \sigma_\parallel^2\equiv\left\langle(\pa-p_0)^2\right\rangle
 =\frac{\kb}{2\kb-3}a_\parallel^2,
 \qquad
 \sigma_{\perp}^2\equiv\langle p_x^2\rangle=\langle p_y^2\rangle
 =\frac{\kb}{2\kb-3}a_\perp^2.
 \label{eq:variances}
\end{equation}

Here, $\sigma_\parallel$ denotes the root-mean-square spread of the longitudinal momentum, while $\sigma_\perp$ denotes the root-mean-square spread of either transverse component, $p_x$ or $p_y$ (i.e., along a single transverse direction). The mean-squared transverse momentum is obtained by adding the contributions from the two transverse components:
\begin{equation}
 \left\langle p_\perp^2\right\rangle
 =\frac{2\kb}{2\kb-3}a_\perp^2.
 \label{eq:perprms}
\end{equation}

Consequently, the root-means-square transverse momentum magnitude is $(\langle p_\perp^2\rangle)^{1/2}=\sigma_\perp \sqrt{2}$. The connection between the $\kb$-type distribution and a Gaussian distribution becomes clear in the limit $\kb\to\infty$. For fixed scale parameters, the distribution approaches
\begin{equation}
 f_{\kb}\longrightarrow f_G=
 \frac{n_b}{\pi^{3/2}a_\perp^2a_\parallel}
 \exp\left[-\frac{\pp^2}{a_\perp^2}
            -\frac{(\pa-p_0)^2}{a_\parallel^2}\right].
 \label{eq:gaussianlimit}
\end{equation}

Thus, the $\kb$-type family approaches an anisotropic Gaussian in momentum space as $\kb$ increases. The scale parameters in Equation~\eqref{eq:gaussianlimit} are $\sigma_\parallel=a_\parallel/\sqrt{2}$ and $\sigma_\perp=a_\perp/\sqrt{2}$. Accordingly, when comparing the $\kb$-type distributions with a Gaussian at fixed root-mean-square momentum spreads, we use the same definitions of $\sigma_\parallel$ and $\sigma_\perp$ as in Equation~\eqref{eq:variances}.
%Thus the chosen \(\kappa\)-type family approaches an anisotropic Gaussian in momentum space. The Gaussian comparison must use the same stated width convention as Equation~\eqref{eq:matchedwidths}.

\subsection{Maxwell--Jüttner Distribution}\label{subsec:mj-distribution}

%A Maxwell--Jüttner distribution is included as a separate relativistic thermal reference for the numerical comparison. 
Unlike the Gaussian distribution obtained in the $\kb\to\infty$ limit of Equation~\eqref{eq:kappadist}, the Maxwell--J\"{u}ttner distribution is defined by an exponential in relativistic energy. It is isotropic in its own rest frame and becomes anisotropic in the plasma frame when it is boosted along the beam direction. To write this distribution in a simplified form, we introduce the dimensionless momentum $\boldsymbol{u}=\bfp/(m_ec)$, with $u_\perp=\pp/(m_ec)$, $u_\parallel=\pa/(m_ec)$, and $\gamma=(1+u_\perp^2+u_\parallel^2)^{1/2}$. We define a dimensionless temperature $\Theta=k_{\mathrm B}T/(m_ec^2)$, where $T$ is the temperature of the beam in its rest frame. The bulk drift Lorentz factor is defined as $\gamma_d=(1-\beta_d^2)^{-1/2}$, where $\beta_d c$ is the drift velocity of the distribution in the plasma frame. The normalized distribution in the plasma frame is \cite{Rafighi:2017ise, Vafin:2018kox}
\begin{equation}
F_{\mathrm{MJ}}(\boldsymbol{u})=
\frac{1}{4\pi\gamma_d\Theta K_2(1/\Theta)}
\exp\!\left[-\frac{\gamma_d(\gamma-\beta_d u_\parallel)}{\Theta}\right],
\label{eq:mj-dist}
\end{equation}
where $K_2$ is the modified Bessel function of the second kind. The dimensionless distribution is normalized such that $\int F_{\mathrm{MJ}}(\boldsymbol{u})\,\dd^3u=1$. The corresponding distribution in physical momentum space is, therefore, $f_{\mathrm{MJ}}(\bfp)=n_b(m_ec)^{-3}F_{\mathrm{MJ}}(\boldsymbol{u})$.

We choose the dimensionless temperature $\Theta$ such that the momentum spread of either transverse momentum component equals $\sigma_\perp$, as defined in Section~\ref{subsec:Gaussian-limit}. The drift velocity is then determined by requiring the mean longitudinal momentum to equal $p_0$. The longitudinal momentum spread is a result of this choice and is not independently matched to that of the $\kappa$-type distribution. The drift velocity is determined by the condition that
\begin{equation}
\frac{p_0}{m_ec}
=\langle u_\parallel\rangle_{\mathrm{MJ}}
=\gamma_d\beta_d\left(\langle\gamma'\rangle_{\mathrm{rest}}+\Theta\right),
\qquad
\langle\gamma'\rangle_{\mathrm{rest}}
=\frac{K_1(1/\Theta)}{K_2(1/\Theta)}+3\Theta,
\label{eq:mj-drift}
\end{equation}
where the subscript ``rest'' denotes an average in the rest frame of the beam.  {The relation follows from the Lorentz transformation of the momentum and the momentum-space integration measure, as detailed in Appendix~\ref{app:mj-moments}. A related discussion of relativistic beam distributions and moments can be found in Wright and Hadley \cite{WrightHadley1975}.%MDPI: please check and revise the citation. %Suman Dey: corrected.
The resulting longitudinal spread is $\sigma_{\parallel,\mathrm{MJ}}=m_ec[\langle u_\parallel^2\rangle_{\mathrm{MJ}}-\langle u_\parallel\rangle_{\mathrm{MJ}}^2]^{1/2}$, where the moments refer to the normalized distribution in Equation~\eqref{eq:mj-dist}. For the fiducial values $p_0/(m_ec)=10$ and $\sigma_\perp/(m_ec)=0.1$, we obtain $\Theta=9.76\times10^{-3}$, $\beta_d=0.99$, and $\gamma_d=9.81$ by reproducing the specified mean longitudinal momentum and the root-mean-square spread of one transverse momentum component. Evaluation of the boosted second moments then yields $\langle u_\parallel\rangle_{\mathrm{MJ}}=10$, $\langle u_\parallel^2\rangle_{\mathrm{MJ}}=100.9$, and $\sigma_{\parallel,\mathrm{MJ}}/(m_ec)\simeq 1.0$. This is approximately twice the reference longitudinal spread $\sigma_\parallel/(m_ec)=0.5$.}

\begin{figure}[H]

    \includegraphics[width=0.85\textwidth]{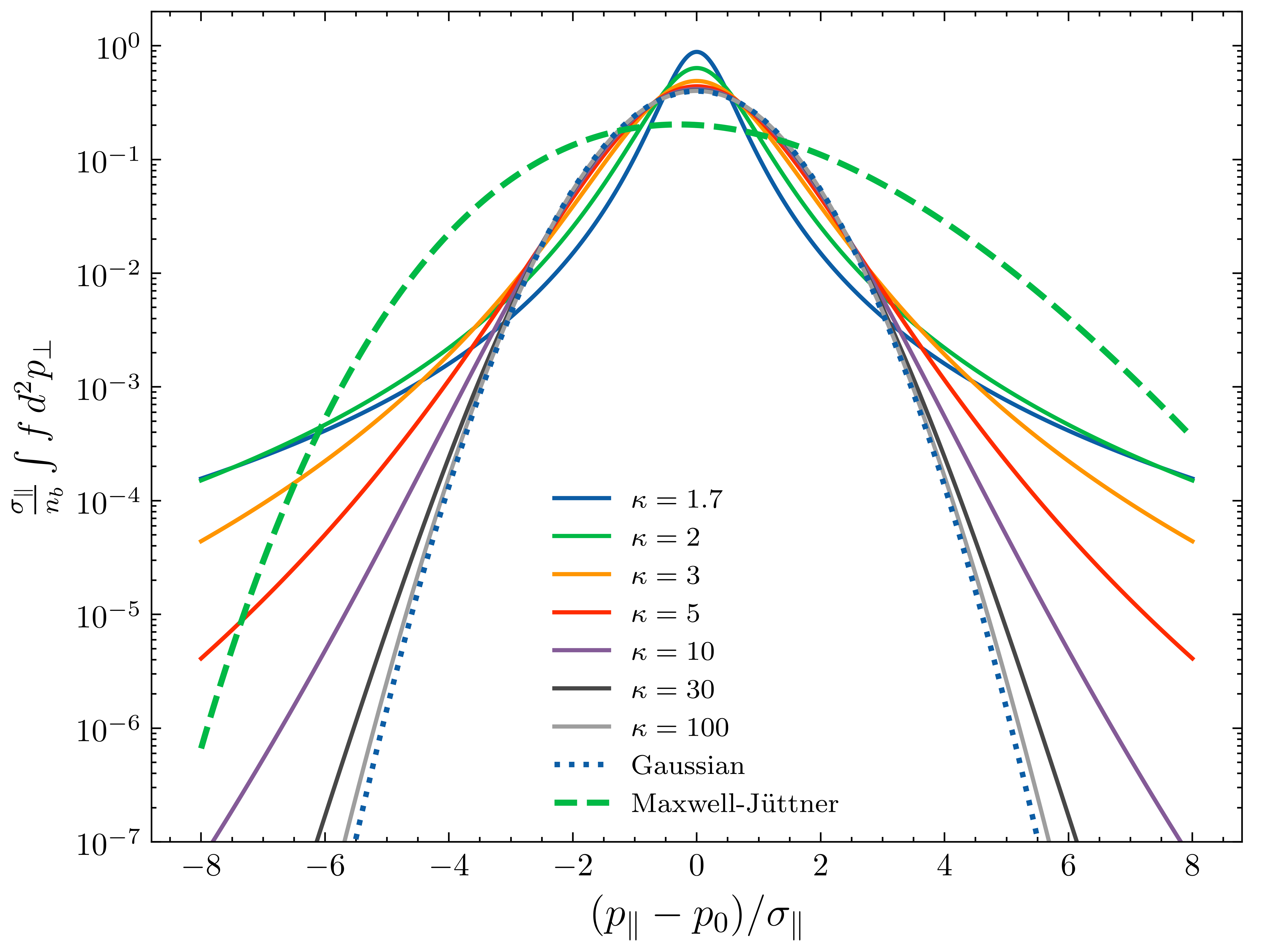}
    \caption{Momentum %MDPI: To avoid large blank spaces on the page, we moved Figure to Section 2.4. Please confirm %Suman Dey: yes, no problem.
 distributions for the $\kb$-type, Gaussian, and Maxwell--J\"{u}ttner beam models. The longitudinal momentum distribution is obtained by integrating over the transverse momentum components and plotted as a function of $(p_\parallel-p_0)/\sigma_\parallel$.}
    \label{fig:dist-functions}
\end{figure}

Figure~\ref{fig:dist-functions} compares the momentum distributions used in the analysis. The solid curves correspond to the $\kb$-type distributions for different values of $\kb$, while the blue dotted and green dashed curves represent the Gaussian and Maxwell--J\"{u}ttner distributions, respectively. Smaller values of $\kb$ produce more pronounced high-momentum tails, whereas the $\kb$-type distribution converges to the Gaussian form in the limit $\kb\rightarrow\infty$.

\subsection{Momentum Derivatives}

To simplify the calculation of the momentum derivatives of the $\kappa$-type distribution in Equation~\eqref{eq:kappadist}, we introduce the auxiliary function
\begin{equation}
 Q_{\kb}=1+\frac{\pp^2}{\kb a_\perp^2}
          +\frac{(\pa-p_0)^2}{\kb a_\parallel^2}.
 \label{eq:Qdef}
\end{equation}

Using the definition in Equation~\eqref{eq:Qdef}, the derivatives of the $\kb$-type distribution in Equation~\eqref{eq:kappadist} with respect to the transverse and longitudinal momenta are given by
\begin{align}
 \frac{\partial f_{\kb}}{\partial p_\perp}
 &=-\frac{2(\kb+1)\pp}{\kb a_\perp^2}\,C_{\kb}Q_{\kb}^{-(\kb+2)},
 \label{eq:dfperp}\\
 \frac{\partial f_{\kb}}{\partial p_\parallel}
 &=-\frac{2(\kb+1)(\pa-p_0)}{\kb a_\parallel^2}\,C_{\kb}Q_{\kb}^{-(\kb+2)}.
 \label{eq:dfpar}
\end{align}

The beam contribution to the electrostatic dispersion relation depends on the projection of the momentum-space gradient along the wavevector. For a gyrophase angle $\phi$, with $p_x=p_\perp\cos\phi$, this projection is
\begin{equation}
 \bfk\cdot\nablaP f_{\kb}
 = k_\parallel\frac{\partial f_{\kb}}{\partial p_\parallel}
 +k_\perp\cos\phi\frac{\partial f_{\kb}}{\partial p_\perp}.
 \label{eq:kgradf}
\end{equation}

To compare the $\kb$-type distribution with a Gaussian distribution, we also require the corresponding momentum derivatives. Differentiating the Gaussian distribution defined in Equation~\eqref{eq:gaussianlimit} gives
\begin{equation}
 \frac{\partial f_G}{\partial p_\perp}
 =-\frac{2p_\perp}{a_\perp^2}f_G,
 \qquad
 \frac{\partial f_G}{\partial p_\parallel}
 =-\frac{2(p_\parallel-p_0)}{a_\parallel^2}f_G.
\label{eq:gaussian-derivatives}
\end{equation}

When the Gaussian and $\kb$-type distributions are compared at fixed root-mean-square momentum spreads, the Gaussian distribution parameters are related to the corresponding width by $a_\parallel=\sqrt{2}\sigma_\parallel$ and $a_\perp=\sqrt{2}\sigma_\perp$. This choice ensures the two distributions have the same longitudinal and transverse widths, so differences in their growth rates reflect differences in shape rather than differences in these widths. These derivatives, together with the Maxwell--J\"{u}ttner derivatives in Equation~\eqref{eq:mj-derivatives}, allow all three distributions to be evaluated using the same resonant growth-rate integral.

For the Maxwell--J\"{u}ttner distribution, the derivatives with respect to dimensionless momentum are given by
\begin{equation}
\frac{\partial F_{\mathrm{MJ}}}{\partial u_\perp}
=-\frac{\gamma_d u_\perp}{\Theta\gamma}F_{\mathrm{MJ}},
\qquad
\frac{\partial F_{\mathrm{MJ}}}{\partial u_\parallel}
=-\frac{\gamma_d}{\Theta}
\left(\frac{u_\parallel}{\gamma}-\beta_d\right)F_{\mathrm{MJ}}.
\label{eq:mj-derivatives}
\end{equation}

Here, $F_{\mathrm{MJ}}$ is normalized over dimensionless momentum space, where $\mathbf{u}=\mathbf{p}/(m_ec)$, as defined in Section~\ref{subsec:mj-distribution}. The Maxwell--J\"{u}ttner distribution thus provides a relativistic thermal reference for comparison with the $\kb$-type and Gaussian distributions. Unlike the other two distributions, whose momentum spreads can be matched directly by adjusting their scale parameters, the Maxwell--J\"{u}ttner distribution is specified by its temperature and drift. In our reference calculation, we choose these parameters to reproduce the prescribed mean longitudinal and transverse momentum spread; the longitudinal spread is then determined by the resulting distribution and need not equal the prescribed value $\sigma_\parallel$.  {In the comparison between the $\kappa$ and Gaussian distributions, the mean longitudinal momentum and the root-mean-square momentum spreads in both the longitudinal and transverse directions are kept equal. However, for the Maxwell--J\"{u}ttner distribution, only the mean longitudinal momentum and the transverse root-mean-square momentum spread are fixed to the same values.} We can substitute the derivatives above into the weak-beam resonant growth-rate integral, derived later in the paper, to compare the electrostatic instability growth rates of different distributions within a consistent framework.

\section{Dispersion Relation for Electrostatic Waves}\label{sec:dispersion}

\subsection{Kinetic Dielectric Function}

For electrostatic perturbations, the electric field is parallel to the wavevector, $\delta\mathbf{E}\parallel\mathbf{k}$. Consequently, the magnetic field perturbation vanishes, $\delta\mathbf{B}=\mathbf{0}$. We consider
a plane-wave dependence of the form
$\exp[i(k_\perp x+k_\parallel z-\omega t)]$, where $x$ and $z$ denote
the transverse and longitudinal coordinates, respectively,
$\omega$ is the angular frequency, and $k_\perp$ and $k_\parallel$
are the transverse and longitudinal components of the wavevector. Under these assumptions, the linear response of the system is described by the relativistic Vlasov--Poisson dispersion relation \cite{fried2015plasma},
\begin{equation}
 \epsilon_L(\omega,\bfk)=1-\frac{\ompe^2}{\omega^2}+\chi_b(\omega,\bfk)=0,
 \label{eq:dispersion}
\end{equation}

The term $-\omega_{pe}^2/\omega^2$ describes the response of the background electrons in the cold-plasma approximation, which neglects their thermal momentum spread and assumes zero equilibrium drift velocity. Although the electrons are initially at rest, they can still respond dynamically to the perturbing electric field. For a beam distribution $f_b(\bfp)$, the susceptibility is given by \cite{bret2010multidimensional, osullivan1973relativistic}
\begin{equation}
 \chi_b(\omega,\bfk)=\frac{e^2}{\epsz k^2}
 \int_{\mathbb{R}^3}
 \frac{\bfk\cdot\nablaP f_b(\bfp)}
 {\omega-\bfk\cdot\bfv}\,\dd^3p.
 \label{eq:chideriv}
\end{equation}

 The denominator describes the wave–particle resonance and is interpreted with the Landau prescription for analytic continuation in $\omega$. For a beam consisting of multiple species with identical momentum distributions, their contributions can be combined by summing their squared charges weighted by their respective number densities.  {To express the susceptibility in terms of the distribution itself, we integrate Equation~\eqref{eq:chideriv} by parts in momentum space. The particle velocity, $\bfv=\bfp/(\gamma m_e)$, must be differentiated when the derivative acts on the resonant denominator. In particular, using Equation~\eqref{eq:dvdpmatrix}, we obtain $k_i k_j(\partial v_i/\partial p_j) = k^2[1-(\widehat{\bfk}\cdot\bfv)^2/c^2]/(\gamma m_e)$. Assuming that the boundary term vanishes in momentum space, the integration by parts yields}
\begin{equation}
 \chi_b(\omega,\bfk)=-\frac{e^2}{\epsz}
 \int_{\mathbb{R}^3} f_b(\bfp)
 \frac{1-(\widehat{\bfk}\cdot\bfv)^2/c^2}
 {\gamma m_e\,[\omega-\bfk\cdot\bfv]^2}\,\dd^3p,
 \label{eq:chiibp}
\end{equation}
where $\widehat{\bfk}=\bfk/k$ is the unit vector along the wavevector. The resonance denominator is treated by analytic continuation in $\omega$, ensuring a consistent evaluation of the momentum integral near the wave--particle resonance. To obtain Equation~\eqref{eq:chiibp}, we use
\begin{equation}
 \frac{\partial v_i}{\partial p_j}=\frac{1}{\gamma m_e}
 \left(\delta_{ij}-\frac{v_i v_j}{c^2}\right)
 \label{eq:dvdpmatrix}
\end{equation}
and assume that the distribution decreases sufficiently rapidly at large momentum for the boundary terms to vanish. Equation~\eqref{eq:chiibp} expresses the beam susceptibility in terms of the distribution function itself, with the relativistic effects explicitly appearing through the Lorentz factor $\gamma$ and the factor $1-(\widehat{\bfk}\cdot\bfv)^2/c^2$. The differences between the Gaussian, $\kb$-type, and Maxwell--J\"{u}ttner distributions are, therefore, captured by their respective forms of $f_b(\bfp)$ in the momentum integral.

\subsection{Growth Rate in the Weak-Beam (or Dilute-Beam) Limit}

For blazar-induced pair beams, the extremely low beam density makes the beam susceptibility a small perturbation to the dielectric response of the background plasma. In this regime, the wave frequency remains approximately equal to the background electron plasma frequency, $\ompe$. We, therefore, write the complex frequency as $\omega=\omega_r+\ii\zeta$, where $\omega_r\simeq\ompe$ and $|\zeta|\ll\omega_r$. Here, $\zeta$ denotes the growth or damping rate, with positive values corresponding to instability. Linearizing Equation~\eqref{eq:dispersion} about the real background Langmuir root and keeping the leading-order beam contribution, we obtain
\begin{equation}
 \zeta(\bfk)\simeq -\frac{\operatorname{Im}\chi_b(\omega_r,\bfk)}
 {\left.\partial_\omega\epsilon_{\mathrm{bg}}(\omega)\right|_{\omega_r}},
 \qquad \epsilon_{\mathrm{bg}}(\omega)=1-\frac{\ompe^2}{\omega^2}.
 \label{eq:weakbeam}
\end{equation}

To evaluate the imaginary part of the susceptibility, we use the Landau identity,
\begin{equation}
 \frac{1}{x+\ii 0^+}=\mathcal{P}\frac{1}{x}-\ii\pi\delta(x),
 \label{eq:landau}
\end{equation}
where $\mathcal{P}$ denotes the Cauchy principal value and $\delta(x)$ is the Dirac delta function. The delta-function term selects particles satisfying the wave--particle resonance condition, $\omega_r=\bfk\cdot\bfv$ \cite{bret2010multidimensional}. Applying this identity to Equation~\eqref{eq:chideriv}, we obtain
\begin{equation}
 \zeta(\bfk)\simeq
 \frac{\pi e^2}{\epsz k^2\,\epsilon_{\mathrm{bg}}'(\omega_r)}
 \int \dd^3p\,\bfk\cdot\nablaP f_b(\bfp)
 \delta\!\left(\omega_r-\bfk\cdot\bfv\right).
 \label{eq:gammageneral}
\end{equation}

The sign of the integral indicates whether the resonant beam particles transfer energy to the wave or absorb energy from it, within the adopted weak-beam (or dilute-beam) approximation and sign conventions. For the cold, stationary background considered here, the Langmuir frequency is $\omega_r\simeq\ompe$, and the derivative of the background dielectric function at this frequency is $\epsilon_{\mathrm{bg}}'(\ompe)=\left.\partial_{\omega} \epsilon_{\mathrm{bg}}(\omega)\right|_{\omega=\ompe} =
2/\ompe$. Substituting this result into Equation~\eqref{eq:gammageneral} gives the growth rate in the weak-beam limit:
\begin{equation}
 \zeta(\bfk)\simeq
 \frac{\pi e^2\ompe}{2\epsz k^2}
 \int \dd^3p\,\bfk\cdot\nablaP f_b(\bfp)
 \delta\!\left(\ompe-\bfk\cdot\bfv\right).
 \label{eq:gammabox}
\end{equation}

A positive value of the resonant integral gives $\zeta>0$, corresponding to exponential growth, whereas a negative value gives damping \cite{nicholson1983introduction, bret2010multidimensional}. This general expression can be used for $f_b=f_{\kb}$, $f_G$, or $f_{\mathrm{MJ}}$, provided the corresponding distribution derivatives are used. It is a weak-beam expression for the growth rate of electrostatic modes driven by blazar-induced pair beams, where the beam density is much smaller than the background plasma density. Therefore, it should not be interpreted as a general expression for the growth rate. If the beam contribution is no longer a small perturbation, the weak-beam expansion is insufficient, and the complex dispersion relation in Equation~\eqref{eq:dispersion} must instead be solved directly.

 {Equation~\eqref{eq:gammabox} describes a resonant instability: the delta function selects the portion of the beam satisfying the Langmuir-wave resonance condition. The resonant gradients determine the instantaneous linear growth or damping of a selected mode, but the magnitude of this growth does not, by itself, determine the eventual transfer of beam energy to the background plasma. In the quasi-linear regime, nearly parallel and oblique resonances can affect beam energy and angular spreading differently~\cite{PerryLyubarsky2021}.}

\subsection{Analytic Integration over the Gyrophase}

For an axisymmetric beam distribution $f_b$, the three-dimensional resonant integral in Equation~\eqref{eq:gammabox} can be reduced to a two-dimensional integral by performing the integration over the gyrophase analytically. At fixed $(\pp,\pa)$, we define
\begin{equation}
 A=\omega_r-k_\parallel v_\parallel,\qquad
 B=k_\perp v_\perp,
 \qquad v_\perp=\frac{\pp}{\gamma m_e},\quad
 v_\parallel=\frac{\pa}{\gamma m_e}.
 \label{eq:AB}
\end{equation}

The resonance condition, $\omega_r-\bfk\cdot\bfv=0$, then becomes $A-B\cos\phi=0$. For $B\ne0$, the integration over the gyrophase can be evaluated using the identity
\begin{equation}
 \int_0^{2\pi}g(\cos\phi)\delta(A-B\cos\phi)\,\dd\phi
 =\frac{2\,g(A/B)}{|B|\sqrt{1-(A/B)^2}},
 \qquad |A|<|B|,
 \label{eq:azimuthidentity}
\end{equation}
where $g$ is a function of $\cos\phi$. The factor of two accounts for the two gyrophase angles that satisfy the resonance condition. When $|A|>|B|$, no such angles exist, and the integral vanishes. Applying this identity to the resonant integral and using the cylindrical momentum-space measure, $\dd^3p=\pp,\dd\pp,\dd\pa,\dd\phi$, we obtain
\begin{align}
 \mathcal{I}_{b}(\bfk)={}&\int \dd^3p\,\bfk\cdot\nablaP f_{b}
 \delta(\omega_r-\bfk\cdot\bfv)\nonumber\\
 ={}&2\int_{0}^{\infty}\pp\,\dd\pp
 \int_{-\infty}^{\infty}\dd\pa\;\mathbf{1}_{|A|<|B|}
 \frac{
 k_\parallel f_\parallel+(k_\perp A/B)f_\perp}
 {|B|\sqrt{1-(A/B)^2}},
 \label{eq:2dintegral}
\end{align}
where $f_\parallel=\partial f_b/ \partial p_\parallel$, $f_\perp=\partial f_b/ \partial p_\perp$, and $\mathbf{1}_{|A|<|B|}$ is an indicator function that equals unity when $|A|<|B|$ and zero when $|A|>|B|$. The derivatives are evaluated at the corresponding momenta $(\pp,\pa)$. This form makes it explicit that only particles satisfying the resonance condition contribute to the growth rate. For the $\kb$-type distribution, the derivatives in Equations~\eqref{eq:dfperp} and \eqref{eq:dfpar} are substituted into Equation~\eqref{eq:2dintegral}. For the Gaussian and Maxwell--J\"{u}ttner distributions, we instead use the corresponding derivatives from Equations~\eqref{eq:gaussian-derivatives} and \eqref{eq:mj-derivatives}. Once the resonant integral has been evaluated for each distribution, the growth rate follows from
\begin{equation}
 \zeta(\bfk)\simeq \frac{\pi e^2}{\epsz k^2\epsilon_{\mathrm{bg}}'(\omega_r)}\mathcal{I}_{b}(\bfk).
 \label{eq:gammareduced}
\end{equation}

The integrand contains an apparent square-root singularity as $|A/B|\to1$, corresponding to the boundary of this domain. Although this singularity is integrable, it requires care in numerical calculations; an appropriate change of variables or a quadrature method designed for endpoint singularities can improve accuracy. In the numerical implementation, we do not integrate this square-root singularity directly. Instead, we solve the resonance condition analytically for the dimensionless longitudinal momentum $u_\parallel=p_\parallel/(m_ec)$ at fixed $r=p_\perp/(m_ec)$ and gyrophase $\phi$. Writing $K_\parallel=k_\parallel c/\omega_{pe}$, $K_\perp=k_\perp c/\omega_{pe}$, and $b=K_\perp r\cos\phi$, the forward resonant root for $K_\parallel>1$ is
%\begingroup
\begin{equation}
 u_{\parallel,\mathrm{res}}=\frac{-K_\parallel b+\sqrt{b^2+(K_\parallel^2-1)(1+r^2)}}{K_\parallel^2-1}.
 \label{eq:resonance-root-numerical}
\end{equation}
%\endgroup

 We perform the integration over longitudinal momentum analytically using the Dirac delta function. The corresponding resonance Jacobian is obtained by differentiating the dimensionless resonance condition with respect to $u_\parallel$, giving $|\partial h/\partial u_\parallel|_{\mathrm{res}}^{-1}=\gamma_{\mathrm{res}}/|K_\parallel-\beta_{\parallel,\mathrm{res}}|$. The remaining integral is computed with tensor-product Gauss--Legendre quadrature over $0\le r\le r_{\max}$ and $0\le\phi\le\pi$ (using the reflection symmetry in gyrophase). For the fiducial parameters, the initial growth-rate calculations are performed with $n_r=180$, $n_\phi=80$, and a radial integration cutoff of $r_{\max}=20\sigma_\perp/(m_ec)=2$. By evaluating the integral along the resonance surface, this approach avoids direct numerical integration over the endpoint singularity in Equation~\eqref{eq:2dintegral}. The convergence of the numerical integration is shown in Appendix~\ref{app:convergence} through variations in the quadrature resolution and radial cutoff at fixed wavevectors, independently of the wavevector-grid convergence analysis. The strictly parallel case, $k_\perp=0$, must be treated separately because $B=0$ and Equation~\eqref{eq:2dintegral} is then inapplicable. In this limit, the resonance condition is independent of the gyrophase, and the integral becomes
\begin{equation}
\mathcal{I}_{b}(k_\perp=0)
=2\pi\int_0^\infty \pp\,\dd\pp
\int_{-\infty}^{\infty}\dd\pa\,
k_\parallel f_\parallel
\delta(\omega_r-k_\parallel v_\parallel).
\end{equation}

Thus, we can evaluate both the oblique and strictly parallel cases consistently, using the appropriate form of the resonant integral in each regime.

\section{Electrostatic Growth Rates for Different Beam Distributions}\label{sec:growth}

\begin{figure}[H]

    \includegraphics[width=1.0\textwidth]{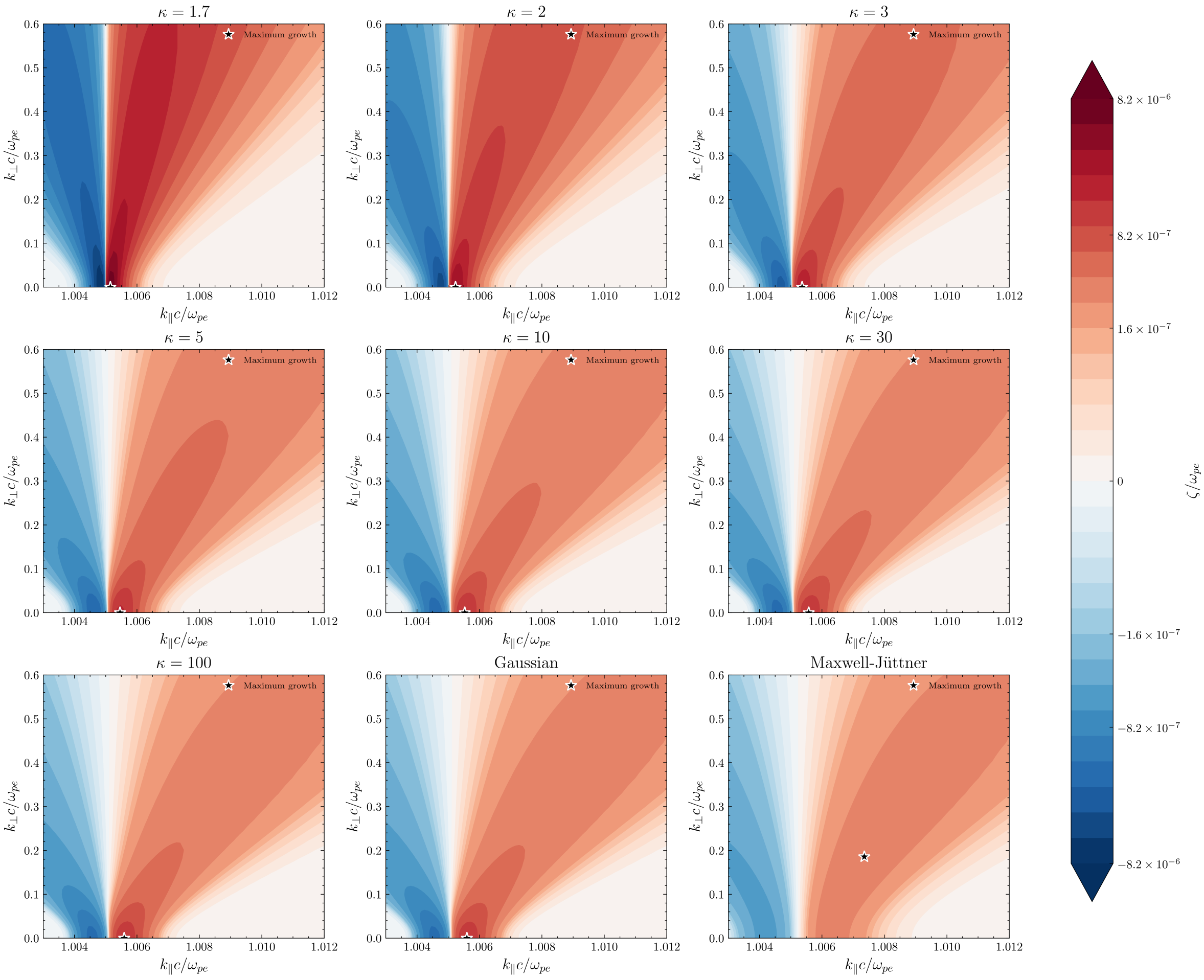}
    \caption{Electrostatic growth-rate maps for the $\kappa$-type distributions with $\kb=1.7$, $2$, $3$, $5$, $10$, $30$, and $100$, together with the Gaussian and Maxwell--J\"{u}ttner distributions. The normalized growth rate $\zeta/\omega_{pe}$ is shown as a function of the normalized longitudinal and transverse wavevector components, $k_\parallel c/\omega_{pe}$ and $k_\perp c/\omega_{pe}$, with red and blue regions indicating growth and damping, respectively. The star markers indicate the maximum growth rates identified on the discrete wavevector grid for each distribution.  {The maximum growth rates are obtained through numerical optimization, and their corresponding wavevector directions are presented in Table~\ref{tab:refined-wavevector-maxima}.}}%MDPI: Please use scientific notations in the figure, e.g., “$8 \times 10^3$”, not “8E3”. Please change the hyphen (-) into a minus sign (−, “U+2212”) in the figure, e.g., “-1” should be “−1”. Please confirm whether the overlapping/incomplete content in this figure affects scientific understanding and if it does, please revise it. %Suman Dey: Corrected.
    \label{fig:gr-2d-maps}
\end{figure}

Figure~\ref{fig:gr-2d-maps} shows the electrostatic growth-rate maps for the $\kappa$-type distributions with $\kb=1.7$, $2$, $3$, $5$, $10$, $30$, and $100$, together with the Gaussian and Maxwell--J\"{u}ttner distributions. For the growth rate calculations, we use $p_0/(m_e c)=10$, $\sigma_\parallel/(m_e c)=0.5$, $\sigma_\perp/(m_e c)=0.1$, and a beam-to-background density ratio $n_b/n_p=10^{-9}$.  {The chosen values define a controlled theoretical test case rather than a self-consistent model of a cosmological blazar-induced pair beam: the characteristic momenta and density ratio are not representative of the full range expected for highly relativistic intergalactic pairs. At fixed normalized distribution and background parameters, Equation~\eqref{eq:gammabox} is linear in $n_b$, so $\zeta/\omega_{pe}\propto n_b/n_p$ in the regime where the weak-beam approximation is valid. Changing the mean momentum or the momentum spreads, in contrast, can also affect the resonant wavevector dependence and the maximizing mode.} The growth rate is shown as a function of the normalized longitudinal and transverse wavevector components, $k_\parallel c/\omega_{pe}$ and $k_\perp c/\omega_{pe}$, respectively. Positive and negative values of $\zeta$ correspond to growth and damping, respectively. The maps show how the momentum-space distribution influences both the magnitude of the growth rate and its dependence on the wavevector. To quantify these differences, we determine the maximum growth rate for each distribution from the numerical wavevector scan,
\begin{equation}
\zeta_{\max,b}
=\max_{k_\parallel,k_\perp}\zeta_b(k_\parallel,k_\perp),
\end{equation}
where $b$ identifies the chosen beam distribution. The corresponding wavevector,
$(k_{\parallel,\max,b},k_{\perp,\max,b})$, identifies the location of the maximum in the scanned parameter space.
 {The propagation angle of a maximizing wavevector is $\theta_{\max,b}=\text{tan}^{-1}(k_{\perp,\max,b}/k_{\parallel,\max,b})$. We optimize the longitudinal and transverse wavevector components in the bounded domain $1.003\leq k_\parallel c/\omega_{pe}\leq1.030$ and $0\leq k_\perp c/\omega_{pe}\leq0.6$, using several initial transverse wavevectors, and compare each result with an independently optimized strictly parallel branch ($k_\perp=0$).} %The results are numerical maxima within the investigated domain, not a mathematical proof of global optimality over all possible wavevectors.}

 To investigate the dependence of the instability growth rate on the mean beam momentum, we perform an additional calculation for $p_0/(m_ec)=10$, $20$, and $50$, while maintaining constant fractional momentum spreads of $\sigma_\parallel/p_0=0.05$ and $\sigma_\perp/p_0=0.01$, as shown in Table~\ref{tab:momentum-sensitivity}. For strictly parallel propagation ($k_\perp=0$), the maximum growth rates of the $\kappa=2$ distribution exceed those of the Gaussian distribution by factors of $2.93$, $3.36$, and $7.38$ for $p_0/(m_ec)=10$, $20$, and $50$, respectively. This indicates that the enhanced growth associated with the $\kappa$ distribution remains significant across the specified momentum range. However, extending the present analysis to cosmological blazar-induced pair beams requires substantially higher Lorentz factors, lower beam densities, and much narrower angular distributions. Under these conditions, both the direction of the fastest-growing electrostatic mode and the relative growth-rate enhancement associated with the $\kappa$ distribution may differ from those obtained for the fiducial parameters considered here. However, we perform an additional calculation in Section~\ref{sec:astrophysical} for a highly relativistic blazar-induced pair beam described by a $\kappa$-type momentum distribution.

\begin{table}[H]
\caption{Mean-momentum sensitivity of the strictly parallel electrostatic branch ($k_\perp=0$), with $\kappa=2$, $\sigma_\parallel/p_0=0.05$, and $\sigma_\perp/p_0=0.01$. For each mean beam momentum, the maximum growth rate is determined within the investigated range of $k_\parallel$.}
\label{tab:momentum-sensitivity}%MDPI: Please cite this table in the text and ensure that the first citation of each table appears in numerical order.  Please add an explanation for the use of bold. If the bold is unnecessary, please remove it. %Suman Dey: corrected and added the reference of the table in the paragraph above.

\begin{tabularx}{\textwidth}{CCCC}
\toprule
$p_0/(m_ec)$ & $\zeta_{\max}^{\kappa=2}/\omega_{pe}$ & $\zeta_{\max,\mathrm{G}}/\omega_{pe}$ & $\zeta_{\max}^{\kappa=2}/\zeta_{\max,\mathrm{G}}$ \\
\midrule
$10$ & $3.65\times10^{-6}$ & $1.24\times10^{-6}$ & $2.93$ \\
$20$ & $5.62\times10^{-6}$ & $1.67\times10^{-6}$ & $3.36$ \\
$50$ & $3.60\times10^{-6}$ & $4.87\times10^{-7}$ & $7.38$ \\
\bottomrule
\end{tabularx}
\end{table}

 {For $p_0/(m_ec)=20$ and $50$, we examine numerical convergence by increasing the Gauss--Legendre quadrature resolution and extending the radial momentum cutoff from $20\sigma_\perp$ to $80\sigma_\perp$. The growth rates vary negligibly under these changes, which indicates the numerical stability of the integration.}
 
  {The results obtained from numerical optimization are presented in Table~\ref{tab:refined-wavevector-maxima}. For all seven $\kappa$ distributions and the Gaussian distribution, the maximum growth rates correspond to strictly parallel propagation ($k_\perp c/\omega_{pe}\sim10^{-9}$). We, therefore, classify these numerical maxima as parallel. The Maxwell--J\"{u}ttner reference instead has an oblique local maximum at $(k_\parallel c/\omega_{pe},k_\perp c/\omega_{pe})=(1.00735,0.186)$, corresponding to $\theta_{\max}=10.45^\circ$. The maximum growth rate for this case is $3.31\times10^{-7}\omega_{pe}$, approximately $6.07\%$ higher than the corresponding value for strictly parallel propagation. However, the Maxwell--J\"{u}ttner distribution also has a different longitudinal root-mean-square momentum spread. Therefore, the difference in the propagation direction of the fastest-growing mode cannot be attributed solely to the shape of the momentum distribution.}
  
   {The maximum growth rates and their corresponding wavevectors are determined through numerical optimization and summarized in Table~\ref{tab:refined-wavevector-maxima}. Figure~\ref{fig:gr-2d-maps} shows the electrostatic growth-rate maps in the $(k_\parallel,k_\perp)$ plane for the different $\kappa$ distributions, the Gaussian distribution, and the Maxwell--J\"{u}ttner distribution. For all $\kappa$ distributions and the Gaussian distribution, the maximum growth rates occur for parallel propagation ($k_\perp=0$). In contrast, the Maxwell--J\"{u}ttner distribution exhibits its maximum at an oblique propagation angle of approximately $10.45^\circ$. The growth rates also decrease as $\kappa$ increases, approaching the Gaussian result.}

\vspace{-6pt}

\begin{table}[H]
\caption{Maximum electrostatic growth rates and corresponding propagation directions for our specified numerical domain. Here, $K_{\parallel,\perp}=k_{\parallel,\perp}c/\omega_{pe}$. The small numerical residuals ($K_\perp\sim10^{-9}$) for the $\kappa$ and Gaussian distributions are consistent with strictly parallel propagation.} %MDPI: Please add an explanation for the use of bold. If the bold is unnecessary, please remove it. % Suman Dey: Corrected
\label{tab:refined-wavevector-maxima}

\small
\begin{tabularx}{\textwidth}{LCCCC}
\toprule
Distribution &
$K_{\parallel,\max}$ &
$K_{\perp,\max}$ &
$\theta_{\max}$ (deg) &
$\zeta_{\max}/\omega_{pe}$ \\
\midrule
$\kappa=1.7$ & 1.005148 & $1.62\times10^{-9}$ &
$9.24\times10^{-8}$ & $7.51\times10^{-6}$ \\

$\kappa=2$ & 1.005233 & $4.13\times10^{-9}$ &
$2.36\times10^{-7}$ & $3.65\times10^{-6}$ \\

$\kappa=3$ & 1.005362 & $3.53\times10^{-9}$ &
$2.01\times10^{-7}$ & $1.99\times10^{-6}$ \\

$\kappa=5$ & 1.005457 & $4.77\times10^{-9}$ &
$2.72\times10^{-7}$ & $1.54\times10^{-6}$ \\

$\kappa=10$ & 1.005526 & $2.68\times10^{-9}$ &
$1.53\times10^{-7}$ & $1.36\times10^{-6}$ \\

$\kappa=30$ & 1.005573 & $2.15\times10^{-9}$ &
$1.23\times10^{-7}$ & $1.27\times10^{-6}$ \\

$\kappa=100$ & 1.005589 & $1.24\times10^{-9}$ &
$7.1\times10^{-8}$ & $1.25\times10^{-6}$ \\

Gaussian & 1.005597 & $1.26\times10^{-9}$ &
$7.15\times10^{-8}$ & $1.24\times10^{-6}$ \\

Maxwell--J\"{u}ttner & 1.007353 & 0.186 &
$1.045\times10^{1}$ & $3.31\times10^{-7}$ \\
\bottomrule
\end{tabularx}
\end{table}

\begin{figure}[H]

    \includegraphics[width=0.85\textwidth]{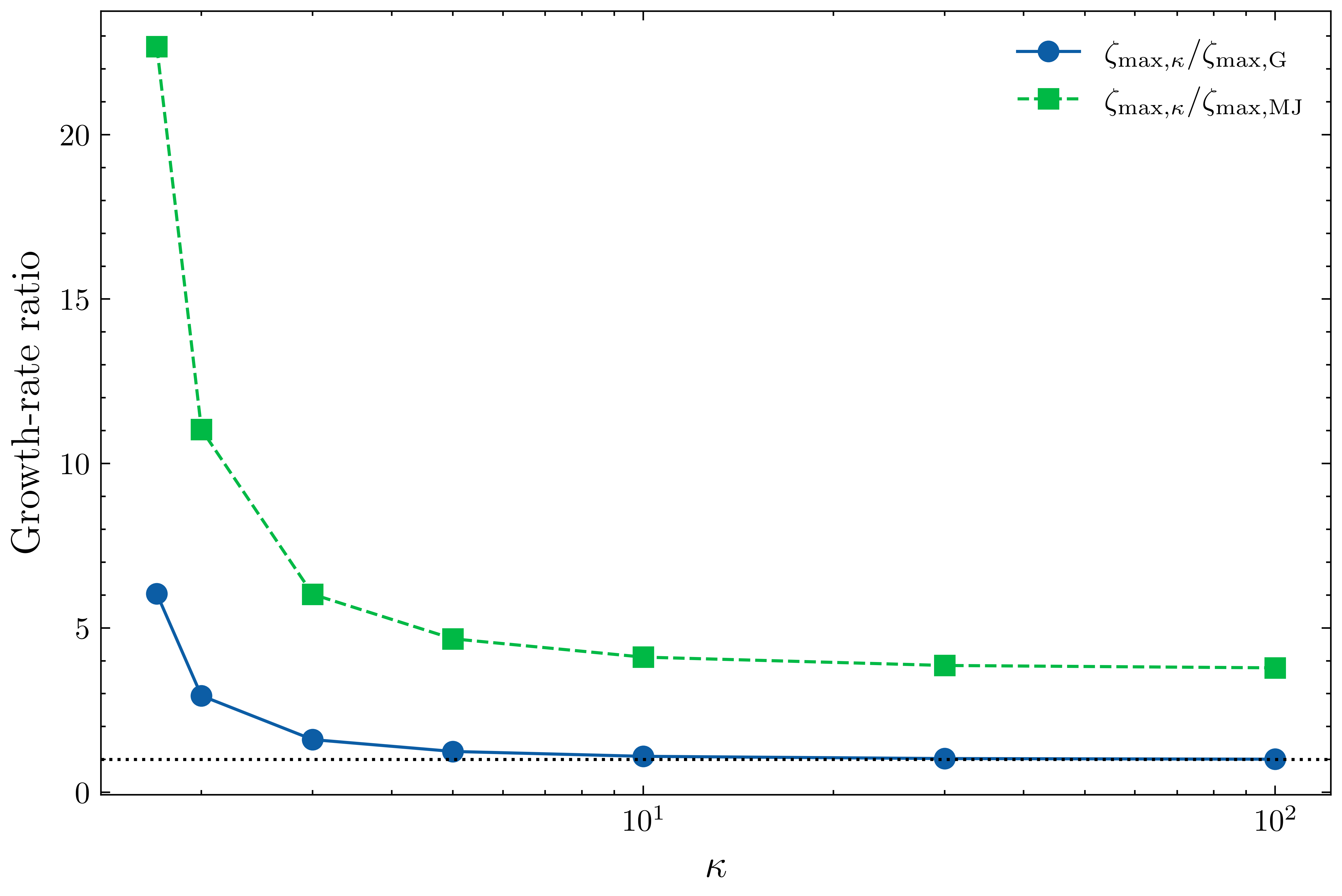}
    \caption{Ratios of the maximum electrostatic growth rates for the $\kb$-type distributions as a function of $\kb$ to those of the Gaussian and Maxwell--J\"{u}ttner distributions. The blue solid curve shows $\zeta_{\max,\kappa}/\zeta_{\max,\mathrm{G}}$, while the green dashed curve shows $\zeta_{\max,\kappa}/\zeta_{\max,\mathrm{MJ}}$. The black dotted line indicates unity; the ratio relative to the Gaussian approaches unity as $\kb$ increases in the $\kb$-type distribution, consistent with the Gaussian limit of the $\kb$-type distribution.}
    \label{fig:gr-rate-ratio}
\end{figure}

Figure~\ref{fig:gr-rate-ratio} compares the maximum growth rates of the $\kb$-type distributions with those of the Gaussian and Maxwell--J\"{u}ttner distributions. The Gaussian-normalized ratio decreases towards unity as $\kb$ increases. This behavior is consistent with the Gaussian limit of the $\kb$-type distribution as $\kb\rightarrow\infty$, provided the corresponding momentum widths are matched. For smaller values of $\kappa$, the more pronounced high-momentum tails are accompanied by higher maximum growth rates in the parameter regime considered here.

The results, therefore, show that the electrostatic instability growth rate depends on the shape of the beam momentum distribution, even when the momentum spreads are fixed. The ratio relative to the Maxwell--J\"{u}ttner distribution also decreases with increasing $\kb$, but approaches a value greater than unity rather than converging to unity. This is consistent with the $\kb$-type distribution approaching the Gaussian distribution, not the Maxwell--J\"{u}ttner distribution.  {The growth-rate ratios between the $\kappa$ and Gaussian distributions show differences in the shapes of their momentum distributions, since their mean longitudinal momenta and both root-mean-square momentum spreads are the same. However, the Maxwell--J\"{u}ttner distribution has a different longitudinal root-mean-square momentum spread. Therefore, the growth-rate ratios between the $\kappa$ and Maxwell--J\"{u}ttner distributions may also be influenced by differences in their longitudinal root-mean-square momentum spreads, in addition to their distribution shapes.} The maxima shown here are obtained from a discrete wavevector grid and should, therefore, be regarded as numerically resolved estimates. Their accuracy is determined by studying the dependence of the results on the wavevector-grid resolution, with the corresponding convergence tests described in Appendix~\ref{app:convergence}.

\section{Astrophysical Implications}
\label{sec:astrophysical}

\begin{figure}[H]

    \includegraphics[width=0.85\textwidth]{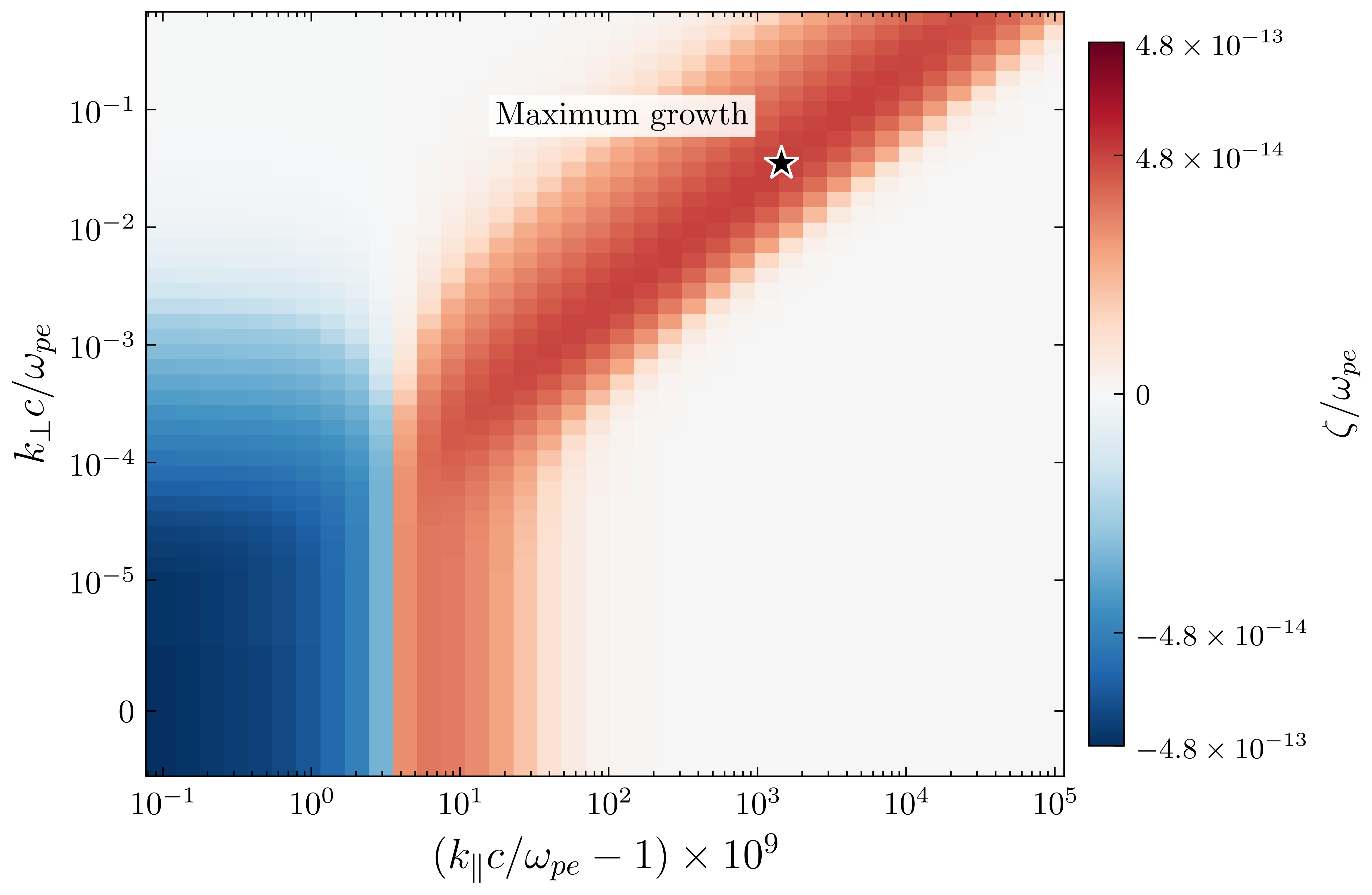}
    \caption{Electrostatic instability growth-rate map for a highly relativistic pair beam with $\kb=2$, $p_0/(m_ec)=10^5$, and $n_b/n_p=5\times10^{-17}$. The star indicates the numerically optimized maximum within the investigated parameter range.} %MDPI: Please use scientific notations in the figure, e.g., “$8 \times 10^3$”, not “8E3”. Please change the hyphen (-) into a minus sign (−, “U+2212”) in the figure, e.g., “-1” should be “−1”. %Suman Dey: Corrected.
    \label{fig:astrophysical-growth-map}
\end{figure}

To study the electrostatic instability under conditions relevant to blazar-induced pair beams, we consider a highly relativistic beam with a $\kb$-type momentum distribution. We adopt $\kb=2$, a mean longitudinal momentum $p_0/(m_ec)=10^5$, and root-mean-square momentum spreads $\sigma_\parallel/(m_ec)=5\times10^4$ and $\sigma_\perp/(m_ec)=10$. The background electron density is considered as $n_p=2\times10^{-7}\,\mathrm{cm}^{-3}$, with a beam-to-background density ratio $n_b/n_p=5\times10^{-17}$ \cite{AlvesBatista:2019ipr, AlawashraPohl2024, Dey:2026nml}. These parameters provide a description of a specific blazar-induced pair beam. For this highly relativistic beam configuration, we adapt the numerical integration to resolve the narrow resonance and investigate the growth rate as a function of the longitudinal and transverse wavevector components. Figure~\ref{fig:astrophysical-growth-map} shows the resulting growth-rate map. Within the investigated wavevector domain, numerical optimization gives a maximum growth rate of $\zeta_{\max}/\omega_{pe}\simeq5.06\times10^{-14}$ at $(k_\parallel c/\omega_{pe}-1)\simeq 1.46\times 10^{-6}$ and $k_\perp c/\omega_{pe}\simeq0.0348$, corresponding to an oblique propagation angle of approximately $2^\circ$.%MDPI: Please confirm if the bold formatting is necessary; if not, please remove it. %Suman Dey: Corrected
The result shows that the $\kb$-type distribution can exhibit growing oblique electrostatic modes at higher beam momenta and lower densities than those considered in Section~\ref{sec:growth}. However, the growth rates and propagation directions depend on the chosen beam parameters. Further studies of the nonlinear evolution are needed to determine whether these instabilities cause significant beam-energy losses~\cite{PerryLyubarsky2021}. %MDPI: Please confirm if the bold formatting is necessary; if not, please remove it. %Suman Dey: corrected.

\section{Conclusions}\label{sec:conclusion}

In this work, we have studied the linear growth of  {electrostatic modes with parallel and oblique propagation} driven by a relativistic pair beam described by a drifted, anisotropic $\kb$-type momentum distribution. The momentum spreads of the $\kb$-type distribution are finite only for $\kb>3/2$. The relation between the scale parameters and these spreads allows us to compare different values of $\kb$ while keeping the momentum spreads fixed. The numerical results show that the beam momentum distribution shape affects the maximum growth rate of  {electrostatic modes}, even when the  {mean momenta and root-mean-square momentum spreads are identical for the $\kappa$ and Gaussian cases}. For the parameter range considered here, the $\kb$-type distributions with more extended high-momentum tails have higher maximum growth rates than the Gaussian distribution. As $\kb$ increases to $\kb=100$, the ratio of the maximum growth rates for the $\kb$-type and Gaussian distributions approaches unity, as expected from the Gaussian limit of the $\kb$-type distribution for large $\kb$. The ratio relative to the Maxwell--J\"{u}ttner distribution also decreases with increasing $\kb$, but approaches a value above unity. This is consistent with the fact that the large-$\kb$ limit is Gaussian rather than Maxwell--J\"{u}ttner. For the Maxwell--J\"{u}ttner distribution, we use the same mean longitudinal momentum and transverse momentum spread as for the other distributions. However, the longitudinal momentum spread is determined by the chosen temperature and drift velocity and may, therefore, be different. The maximum growth rate identifies the fastest-growing electrostatic mode, while the corresponding wavevector determines its wavelength and propagation angle relative to the beam direction.  {For the parameters considered in this study, the maximum growth rates occur for parallel propagation in the $\kappa$ and Gaussian distributions, whereas the Maxwell--J\"{u}ttner distribution exhibits its maximum at an oblique angle of approximately $10.45^\circ$.} 

This result is important for blazar-induced pair beams, which are nonthermal momentum distributions with finite spreads both parallel and perpendicular to the beam direction. Moreover, establishing a realistic analytical form for the beam distribution remains challenging, motivating the investigation of alternative distribution functions. Comparing the maximum growth rates and their corresponding wavevectors, therefore, determines how the pair beam distribution function influences the linear development of plasma instabilities. We can compare the maximum growth rate with the inverse-Compton cooling rate to assess whether the instability has enough time to grow during beam propagation. However, the maximum linear growth rate alone does not determine the eventual energy loss of the beam, which also depends on nonlinear saturation and the evolution of the excited plasma waves.

The astrophysical example presented in Section~\ref{sec:astrophysical} demonstrates that highly relativistic beams with a $\kb$-type momentum distribution can exhibit growing oblique electrostatic modes. However, the growth rates and preferred propagation directions depend on the beam parameters and momentum distribution. Further investigation of nonlinear plasma processes and electromagnetic cascade propagation is necessary to determine whether these instabilities significantly contribute to the energy losses of blazar-induced pair beams. %MDPI: Please confirm if the bold formatting is necessary; if not, please remove it. %Suman Dey: Corrected.

The present analysis has some limitations. We assume the background electrons are cold and stationary. Including a finite background-electron temperature would require a kinetic description of their response and could introduce Landau damping, thereby modifying the growth rates \cite{Dey:2025wdj, BeckEtAl2023}. This may also be interesting when considering additional diffuse electron populations associated with MeV cosmic rays \cite{Alawashra:2025tcj, Yang:2023pwq}. These results provide a basis for further numerical studies of how non-Maxwellian momentum distributions affect beam–plasma instabilities in the linear regime. Future work should focus on the nonlinear evolution of instabilities driven by such beams.

\vspace{6pt}

\funding{This research received no external funding.}
%\institutionalreview{Not applicable.}
%\informedconsent{Not applicable.}
\dataavailability{Data available upon request.}
\acknowledgments{I gratefully acknowledge the computing support provided by Technische Universit\"{a}t Braunschweig during the course of this work. I am grateful to the reviewer for the constructive input. I would also like to thank the developers of \texttt{SciencePlots} \cite{SciencePlots} for making scientific figures look so much more beautiful. Their work greatly helped us prepare the figures for this paper, and we appreciate it.} %MDPI: Attention Assigned Editor: please confirm whether the funding information in the Acknowledgments section should be moved to the Funding section. %Suman Dey: Checked and modified.
\conflictsofinterest{The author declares no conflicts of interest.}

\appendixtitles{yes} % Leave argument "no" if all appendix headings stay EMPTY (then no dot is printed after "Appendix A"). If the appendix sections contain a heading then change the argument to "yes".
\appendixstart
\appendix
\section[\appendixname~\thesection]{Numerical Convergence}\label{app:convergence}
\begin{figure}[H]

    \includegraphics[width=0.9\textwidth]{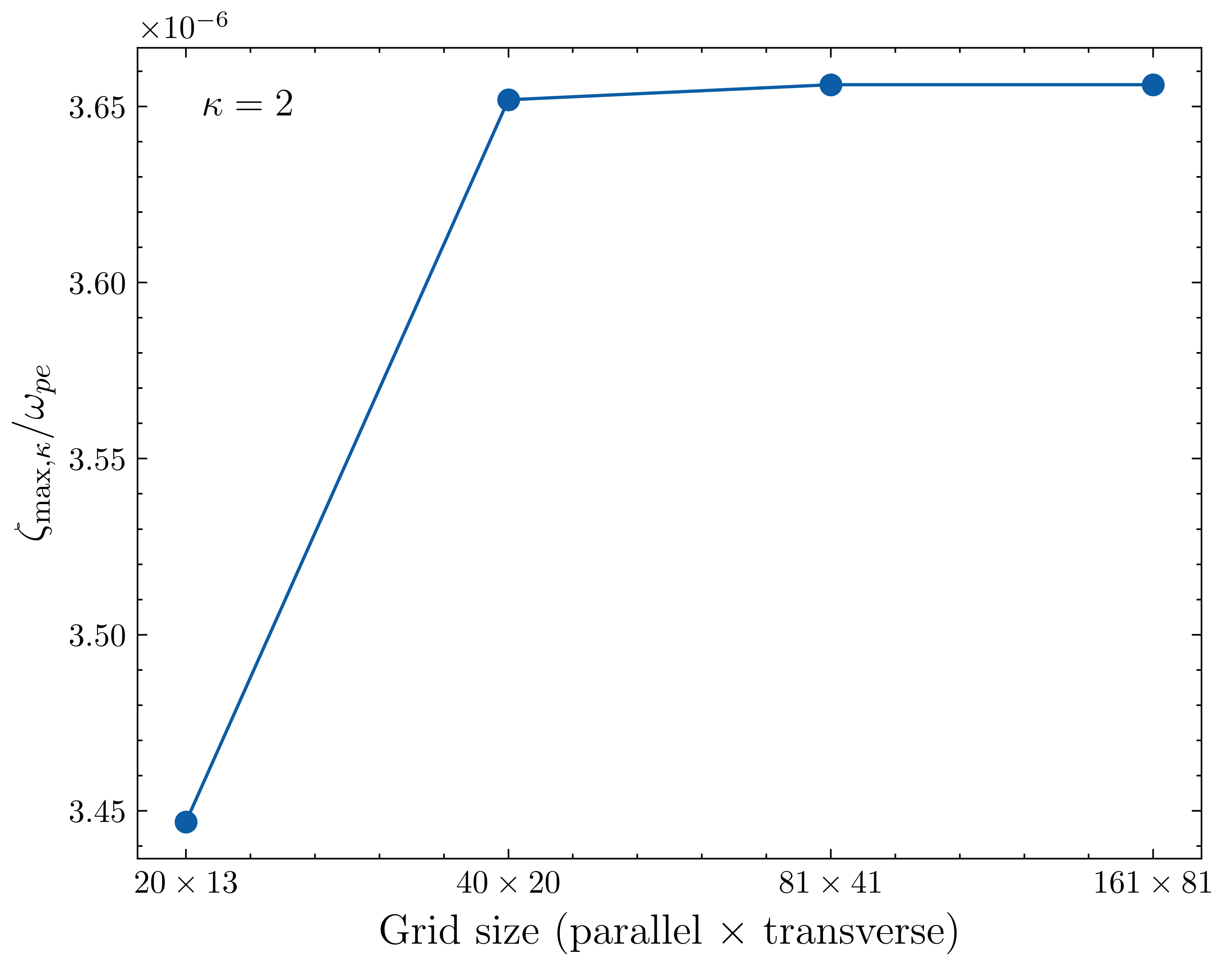}
    \caption{Convergence of the maximum electrostatic growth rate with wavevector-grid resolution for $\kb=2$. The grid size is given as the number of parallel and transverse wavevector points. The results converge as the grid is refined, with the growth rate becoming similar between the $81\times41$ and $161\times81$ grids.}
    \label{fig:gr-rate-convergence}
\end{figure}

%\subsection[\appendixname~\thesubsection]{}
We test the convergence of the maximum growth rate by increasing the resolution of the wavevector grid. Figure~\ref{fig:gr-rate-convergence} shows the result for $\kb=2$. The maximum growth rate increases noticeably when the grid is refined from $20\times13$ to $40\times20$, indicating that the coarsest grid does not fully resolve the maximum. Further refinement to $81\times41$ produces only a small change of approximately $0.1\%$, while the $161\times81$ grid gives the same result to the reported precision. This demonstrates that the maximum growth rate is well converged at a grid resolution of $81\times41$ for the case considered here.

 {To examine numerical convergence, we calculate the growth rates at the maximum-growth wavevectors obtained from the original grid, keeping the wavevectors fixed. Increasing the Gauss--Legendre quadrature orders from $(n_r,n_\phi)=(180,80)$ to $(360,240)$ produces negligible changes in the calculated growth rates. These results indicate the numerical stability of the momentum integration at the selected wavevectors, although they do not establish convergence across the entire wavevector domain.}

 {We also examine the convergence of the radial momentum integration for $\kappa=1.7$, $\kappa=2$, the Gaussian distribution, and the Maxwell--J\"{u}ttner distribution at their respective maximum-growth wavevectors. The radial cutoff, is defined by $r=p_\perp/(m_ec)$, which  increases from $r_{\max}=2$ to $4$ and $8$, with corresponding radial quadrature orders of $180$, $360$, and $720$, while keeping  angular quadrature is fixed at $n_\phi=160$. The growth rates remain nearly unchanged as the radial cutoff increases, which indicates the numerical stability of the integration at the selected wavevectors.}

\begin{table}[H]
\caption{Resonant momentum-integration convergence at selected discrete-grid maxima. Growth rates are normalized by $\omega_{pe}$. The quadrature-order comparison is between $(180,80)$ and $(360,240)$ at $r_{\max}=2$; the radial-cutoff comparison is between $r_{\max}=2$ and $8$ with the radial order scaled accordingly.} %MDPI: Please add an explanation for the use of bold. If the bold is unnecessary, please remove it. %Suman Dey: Corrected.
\label{tab:momentum-convergence}

\begin{tabularx}{\textwidth}{LCC}
\toprule
Distribution & $\zeta/\omega_{pe}$ at $r_{\max}=2$ & Radial Cutoff Difference (\%)\\
\midrule
$\kappa=1.7$ & $7.4\times10^{-6}$ & $<10^{-4}$\\
$\kappa=2$ & $3.6\times10^{-6}$ & $<10^{-4}$\\
Gaussian & $1.2\times10^{-6}$ & $\ll10^{-4}$\\
Maxwell--J\"{u}ttner & $3.3\times10^{-7}$ & $\ll10^{-4}$\\
\bottomrule
\end{tabularx}
\end{table}

We further examine numerical consistency by comparing two formulations of the resonant momentum integral, one using a simplified Jacobian and the other using the Jacobian obtained directly from the delta-function integration. The comparison is performed at the selected wavevectors using $r_{\max}=4$ and $(n_r,n_\phi)=(360,240)$. Both formulations yield nearly identical growth rates, which indicate the numerical consistency of the calculations.

\section[\appendixname~\thesection]{Mean Momentum and Momentum Spreads of the Maxwell--J\"{u}ttner Distribution}\label{app:mj-moments}
 {We outline the moment relation used in Equation~\eqref{eq:mj-drift} for the boosted Maxwell--J\"{u}ttner distribution. We use primed quantities to represent the isotropic beam rest frame, in which $\gamma'=\sqrt{1+|\boldsymbol{u}'|^2}$ and the distribution function is given by $F'(\boldsymbol{u}')=\exp(-\gamma'/\Theta)/[4\pi\Theta K_2(1/\Theta)]$. Applying a Lorentz boost along the beam direction with velocity $\beta_d c$, we obtain}
{\begin{equation}
u_\parallel=\gamma_d(u'_\parallel+\beta_d\gamma'),\qquad
\gamma=\gamma_d(\gamma'+\beta_d u'_\parallel),\qquad
\dd^3u=\frac{\gamma}{\gamma'}\,\dd^3u'.
\label{eq:appendix-mj-transform}
\end{equation}

The factor $1/\gamma_d$ in Equation~\eqref{eq:mj-dist} defines the normalization. For an arbitrary function $g$ of the particle momentum, its expectation value can be written as
{\begin{equation}
 \langle g\rangle_{\mathrm{MJ}}=\int\dd^3u'\,F'(\boldsymbol{u}')
 \left(1+\beta_d\frac{u'_\parallel}{\gamma'}\right)
 g\!\left(\boldsymbol{u}(\boldsymbol{u}')\right).
 \label{eq:appendix-mj-average}
\end{equation}

Isotropy makes odd rest-frame momentum averages vanish. Applying integration by parts to the normalization integral of the rest-frame Maxwell--J\"{u}ttner distribution, we obtain the identity $\langle {u'_\parallel}^2/\gamma'\rangle_{\rm rest}=\Theta$. Therefore,
\begin{align}
 \langle u_\parallel\rangle_{\mathrm{MJ}}
&=\gamma_d\beta_d\left(\langle\gamma'\rangle_{\mathrm{rest}}+\Theta\right),\\
 \langle\gamma'\rangle_{\mathrm{rest}}
 &=\frac{K_1(1/\Theta)}{K_2(1/\Theta)}+3\Theta.
 \label{eq:appendix-mj-firstmoment}
\end{align}

The same transformation provides the second moments needed to evaluate the longitudinal spread without introducing a new distribution prescription:
\begin{align}
 \langle u_\parallel^2\rangle_{\mathrm{MJ}}
 &=\gamma_d^2\left[(1+2\beta_d^2)\langle {u'_\parallel}^2\rangle_{\mathrm{rest}}+\beta_d^2\langle {\gamma'}^2\rangle_{\mathrm{rest}}\right],\\
 \langle u_x^2\rangle_{\mathrm{MJ}}
 &=\langle {u'_x}^2\rangle_{\mathrm{rest}}
 =\frac{\langle {\gamma'}^2\rangle_{\mathrm{rest}}-1}{3},\\
 \langle {\gamma'}^2\rangle_{\mathrm{rest}}
 &=\frac{\displaystyle\int_1^\infty\!\dd\gamma'\,\gamma'^3\sqrt{{\gamma'}^2-1}\,\exp(-\gamma'/\Theta)}
 {\displaystyle\int_1^\infty\!\dd\gamma'\,\gamma'\sqrt{{\gamma'}^2-1}\,\exp(-\gamma'/\Theta)}.
 \label{eq:appendix-mj-secondmoments}
\end{align}

The drift velocity and temperature are obtained by reproducing the specified mean longitudinal momentum, $\langle u_\parallel\rangle_{\mathrm{MJ}}=p_0/(m_ec)$, and transverse momentum spread, $[\langle u_x^2\rangle_{\mathrm{MJ}}]^{1/2}=\sigma_\perp/(m_ec)$. The resulting parameters then determine the longitudinal momentum spread $\sigma_{\parallel,\mathrm{MJ}}$.

%\begin{adjustwidth}{-\extralength}{0cm}
%\centering %% If there is a figure in wide page, please release command \centering, for Table, ``\textwidth" should be ``\fulllength"

\reftitle{References}
%\reftitle{References}
\bibliography{ref}

%\end{adjustwidth}
\end{document}